\documentclass[12pt,preprint]{aastex}
 
\usepackage{epstopdf}
\usepackage{graphics,graphicx}
 
\shorttitle{Closure}
\shortauthors{Carilli}
 
\begin{document}

\title{HERA Mock Observations: Looking for Closure\\
HERA Memorandum Number 13 \\ 
May 13, 2016 
}

\author{C.L. Carilli\altaffilmark{1,2}, P. Sims\altaffilmark{2}}

\email{ccarilli@aoc.nrao.edu}

\altaffiltext{1}{National Radio Astronomy Observatory, P. O. Box 0,
Socorro, NM 87801}
\altaffiltext{2}{Cavendish laboratory, Cambridge University, UK}

\begin{abstract}
\noindent 

We investigate the use of closure phase as a method to detect the HI
21cm signal from the neutral IGM during cosmic reionzation. Closure
quantities have the unique advantage of being independent of
antenna-based calibration terms. We employ realistic, large area sky
models from Sims et al. (2016). These include an estimate of the HI
21cm signal generated using 21cm FAST, plus continuum models of both
the diffuse Galactic synchrotron emission and the extragalactic point
sources. We employ the CASA simulator and adopt the Dillon-Parsons
HERA configuration to generate a uv measurement set. We then use AIPS
to calculate the closure phases as a function of frequency ('closure
spectra'), and python scripts for subsequent analysis.  We find that
the closure spectra for the HI signal show dramatic structure in
frequency, and based on thermal noise alone, the redundant HERA-331
array should detect these fluctuations easily. Comparatively, the
frequency structure in the continuum closure spectra is much smoother
than that seen in the HI closure spectra.  Unfortunately, when the
line and continuum signals are combined, the continuum dominates the
visibilities at the level of $10^3$ to $10^4$, and the line signal is
lost.  We have investigated fitting and removing smooth curves in
frequency to the line plus continuum closure spectra, and find that
the continuum itself shows enough structure in frequency in the
closure spectra to preclude separation of the continuum and line based
on such a process.  We have also considered the subtraction of the
continuum from the visibilities using a sky model, prior to
calculation of the closure spectra. We find that if 99\% of the
continuum can be subtracted from the visibilities, then the line
signal can be seen in the residuals after subsequent smooth curve
fitting and removal, although the advantages of such an approach are
not clear at this point.

\end{abstract}

\clearpage
\newpage

\section{Introduction}

Detecting the HI 21cm signal from the neutral intergalactic medium
during cosmic reionization, and into the preceding dark ages, has been
one of the paramount goals of modern astrophysics for the last decade
(Morales \& Wyithe 2010). However, this task is complicated by the
much stronger foreground continuum emission. 

A powerful distinguishing property of the foregrounds is that they are
dominated by spectrally smooth emission. This is in stark contrast
with the 21-cm emission which is expected to fluctuate rapidly in both
its spatial and spectral dimensions. A naive solution\footnote{
Without a priori knowledge of the covariance between the foregrounds
and the 21-cm signal in the data, independent subtraction of a
foreground model from the data prior to estimation of the quantity of
interest will produce biased estimates of said quantity. As
such, joint estimation of the foregrounds and 21-cm signal is
essential for obtaining statistically robust estimates of the 21-cm
signal (Sims et al. 2016).}, therefore, would be to attempt to removed
the foregrounds in the spectral domain by fitting a smoothly varying
function (such as a polynomial) in frequency to either the
visibilities or the spectral image cubes.  However, in his PhD thesis
work, A. Datta showed that the chromatic response of an interferometer
for very wide field imaging imprints a spectral signature on the
visibility data which is impossible to remove using standard continuum
subtraction techniques via smooth curve fitting to the visibilty
spectra, such as UVLIN in AIPS or uvcontsub in CASA, or point-by-point
smooth curve fitting to a spectral image cube.  The continuum can
still be removed properly, in theory, through a frequency dependent
subtraction from the visibilities of an accurate continuum model
generated from the data themselves. However, Datta et al. showed that
such a subtraction requires remarkably accurate complex gain
calibration as a function of frequency (0.1\%; Datta et al. 2009;
2010).

These facts have led to consideration of alternative methods for
detecting the HI 21cm signal, through 'foreground avoidance' in delay
spectra (Parsons et al. 2012; Morales et al. 2012). The method
involves separating HI from continuum in the line of sight
vs. sky-plane power-spectral space. In this space, the maximum wave
number (or spectral frequency) for flat-spectrum continuum emission
due to the chromatic response of a given interferometric baseline is
set by the maximum delay of the baseline for sources at the
horizon. Hence, the line signal in the line-of-sight direction
(frequency) emerges from 'the wedge' of continuum emission at large
wave number (eg. Datta et al. 2010). This avoidance method still
requires tight control of the spectral response of other parts of the
array, such as the antennas and data transmission system, and/or very
accurate calibration of the spectral response (bandpass) with time, to
avoid coupling the continuum signal to the line, and hence causing
'bleeding' of the continuum signal into the EoR window (eg. Pober et
al. 2016).

In this memo, we consider an alternate approach for discovering the HI
signal using the closure phases of the interferometer. Closure phase
results from a simple product of the three visibility pairs from three
antennas (Jennison 1958).  It was recognized early in the field of
radio interferometry that closure quantities are independent of
antenna-based phase and amplitude calibration terms.  Hence, to the
degree that array calibration is separable into antenna-based terms,
closure phases are independent of calibration and calibration errors,
ie. close quantities are a robust 'observable' of the true sky
signal. This fact was used in early radio interferometry, and in
particular, VLBI, when maintaining phase coherence was
problematic. Closure quantities are still used extensively in optical
interferometry, as well as being the primary diagnostic for
antenna-based calibration errors in phase-connected radio
interferometers (Perley 1999).

Note that in this memo, the goal is not to characterize the HI 21cm
signal from reionization, nor to consider the physical interpretation
and its implications for the physics of reionization.  These are early
days in HI 21cm cosmology, when mere detection of the signal remains
paramount.  Given the robust nature of closure phase to antenna-based
calibration, herein we consider the simple questions: is the HI 21cm
line signal from reionization obvious in the closure phase behaviour
as a function of frequency? Does the behaviour of the closure phases
due to the line signal as a function of frequency differ substantially
from that of the continuum?  And are the two separable in a simple
way?
 
\section{Closure phase}

We briefly review the definition of closure phase (see Cornwell \&
Fomalont 1999). 

The van Cittert-Zernike theorem states that the time averaged
cross-correlation of the electric field voltages measured at two
distinct points in space ($i,j$), which we will call the 'true sky'
visibility, $V_{i,j}^s$, is the Fourier transform of the sky
brightness distribution:

$$ V_{i,j,\nu}^s (u,v) = \int_l \int_m I_\nu(l,m) e^{-2\pi i (ul +
vm)} dl dm $$

\noindent where the small angle approximation has been assume, with
$u$ and $v$ being orthogonal baseline lengths in units of wavelengths,
and $l$ and $m$ as small angle coordinates (direction cosines),
measured with respect to the $u$ and $v$ axes. The subscript $\nu$
denotes measurements as a function of frequency, which we will omit
heretofore. This quantity is separable into a complex expression
of the form:

$$ V_{i,j}^s (u,v) = A^s_{i,j} e^{i\phi_{i,j}^s} $$

\noindent where $A^s_{i,j}$ is the true sky visibility amplitude (ie.
the amplitude of the sinusoidal fringe on the sky), and $\phi_{i,j}^s$
is the phase of the complex visibility (the position of the sinusoidal
fringe). 

A real interferometer will introduce both thermal noise and complex
gain terms, $G$ (ie. amplitude and phase terms due to the instrument
response), that will alter the sky visibility to a measured quantity,
$V_{i,j}^m (u,v)$:

$$ V_{i,j}^m = G_i G_j^\ast V_{i,j}^s = a_i e^{i\theta_i} a_j e^{-i\theta_j}
A^s_{i,j} e^{i\phi_{i,j}^s} ~~ {\rm +~ noise} $$

\noindent where $\theta_i$ is the phase introduced to the visibility
by the antenna electronics or optics, and $a_i$ is the gain amplitude
of the antenna plus electronics. This assumes that the complex gain
on a given visibility is separable to antenna-based terms.

From this, we can see that the resulting measured visibility phase is
the sum of exponents:

$$ \phi_{i,j}^m = \phi_{i,j}^s + (\theta_i - \theta_j) ~~ {\rm +~ noise} $$

The 'bi-spectrum' or 'triple product' for an interferometric measurement
is defined as:

$$C_{i,j,k}^m = V_{i,j}^m V_{j,k}^m V_{k,i}^m$$

\noindent It is easy to see from the equations above that the phase of
this complex measurement, or closure phase, is, again, the sum of
exponents:

$$ \phi_{i,j,k}^m = \phi_{i,j}^s + (\theta_i - \theta_j) +
\phi_{j,k}^s + (\theta_j - \theta_k) + \phi_{k,i}^s + (\theta_k -
\theta_i) ~~ {\rm +~ noise} $$

\noindent The antenna based phase terms then cancel in such a
triangle, leading to:

$$  \phi_{i,j,k}^m = \phi_{i,j,k}^s ~~ {\rm + ~noise}$$

\noindent The implication is that the measured closure phase is {\sl
independent of antenna-based calibration terms}, and represents a
direct measurement of the true closure phase due to structure on the
sky. This fact was recognized early in the field of astronomical
interferometry (Jennison 1958), and is still often used in situations
where instrumental phase stability and determination of antenna-based
calibration terms may be difficult, such as in certain VLBI
applications (historically), and optical interferometry (see Thomson,
Moran, \& Swenson 2007). Note that there is a analogous 'closure amplitude'
based on the combination of four visibility measurements (Cornwell
\& Fomalont 1999).

Of course, this conclusion relies on the assumption that the phase
induced by the system is factorizable into antenna based terms,
ie. that the correlator or other aspects of the system do not introduce
phase terms that depend on the particular cross correlation for a
visibility.  Such non-closuing terms are known as 'closure errors',
and remain an important diagnostic on the quality of antenna-based
calibration in interferometers (Fomalont \& Perley 1999).

\section{Sky models, Mock Observations, and Generation of Closure Phase}

The HI 21cm signal was generated using 21cmFAST (Mesinger et al.
2011).  In order to investigate very wide field effects, the
input model is 45$^o$ across, generated by tiling a series of line
cubes with structure based on the excursion set analysis for the IGM
during cosmic reionization (Sims et al. 2016). 
A very wide-band cube was generated, from roughly
100MHz to 200MHz, with 0.21MHz channels. From this, we selected an
8MHz band at $z = 10$, or 130MHz for the HI line, and for which the
IGM in the model has a mean neutral fraction of 0.5.

The continuum model corresponds to a higher Galactic latiture field
($b=-78^o$). This model is also very wide field (45$^o$), and includes
models for both the Extragalactic point sources, and the Galactic
diffuse synchrotron emission based on existing low frequency
observations but extrapolated to smaller scales in a statistically
defensible manner (Sims et al. 2016).

The input sky model is in FITS image cube format, in units of Jy
pixel$^{-1}$, with a pixel size of $85"$ (converted from the original
Kelvin brightness unit). Images resulting from these models are
presented in Carilli \& Sims (2016). The model is folded through the
CASA simulator, SIMOBSERVE, using the Dillon-Parsons (2016) 'split
core' array for HERA-350, although we only use the shortest baselines
for the analysis herein.  We then export the visibility measurement
set in FITS format, and load into AIPS. Closure phases as a function
of frequency ('closure spectra'), are then generated using the AIPS
task CLPLT, in ascii format. Further analysis is done in Python.

The hexagonal configuration has a distinct advantage for closure phase
consideration due to the many redundant triangles. We assume the
number of independent triangles based on the Dillon \& Parsons
HERA-331 'split core' configuration. In this memo, we consider only
the three shortest equilateral triangles in the array, ie. triangles of
1, 2, and 3 times the grid spacing of 14.6m. There are roughly 75 of
each triangle in the array (although there may be a better way of
combining baselines in the configurtation that improves signal to
noise).

Under these assumptions, we calculate the signal-to-noise in the
measurements based on thermal noise. We assume a total integration
time of 100 hours. The transit time for the 10$^o$ primary beam is
40min, so this would take 150days.  We adopt a channel width of
0.21MHz and system performance as per de Boer et al. (2016), leading
to an rms per visibility of 12mJy. The typical HI signal amplitude on
the short baselines $\sim 4$mJy.  Hence, the S/N per visibility phase
measurement is $\sim 0.33$. This improves by a factor root(3/2) for
the closure triangle (I think), implying a signal-to-noise for each
closure triangle of $\sim 0.40$. Summing the 75 redundant triangles
then leads to a signal-to-noise of $\sim 3.5$ per closure triangle
length. Adopting the analysis of Wrobel \& Walker (1999) for phase
noise as a function of signal-to-noise, leads to a phase error of
$\sigma \sim 17^o$ per channel in the closure spectra. 

\section{Results}

We consider the HI and continuum closure spectra separately, in order
to investigate their relative behaviour.  We then consider the closure
spectra for the summed line and continuum emission, and methods to
extract the line signature in the presence of the much stronger
continuum.

\subsection{Line signal only}

Figure 1 shows the frequency dependence of the closure phase for the
three different triangles. The line signal shows large fluctuations
across the spectrum.  This is consistent with the large changes in sky
signal across the band, ie. substantially changing structure with
frequency on scales of the bandwidth and spatial resolution of the
baseline. Note that there is a $\pm 180^o$ ambiguity inherent in phase
measurements, leading to apparent 'jumps' at high positive or negative
values. Some of these have been rectified as logic dictates.

\begin{figure}[ht]
\epsscale{0.47}
\plotone{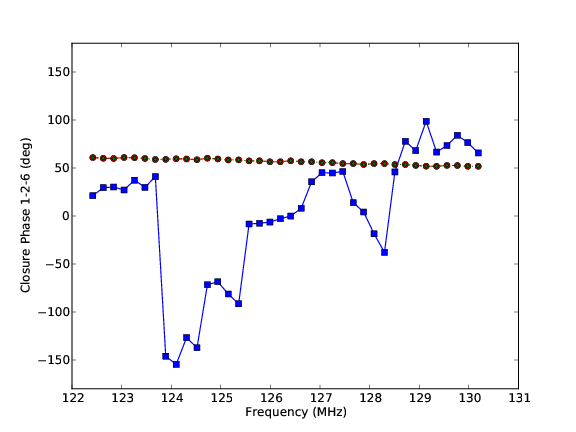}
\plotone{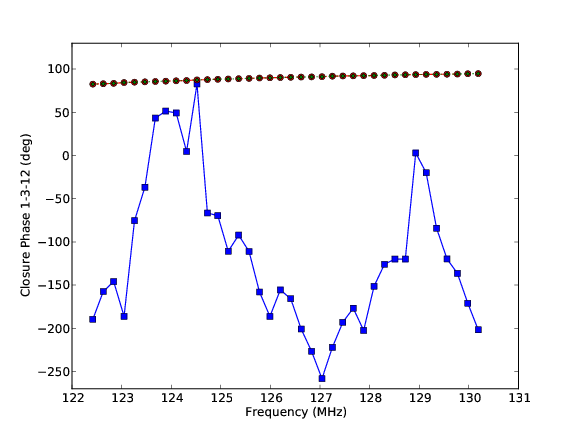}
\plotone{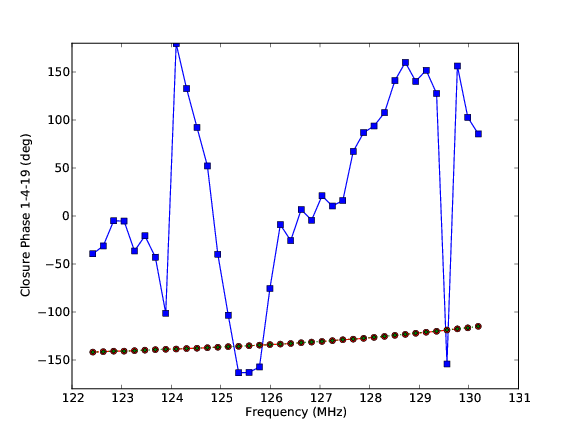}
\caption{\small The closure phase spectra for the three smallest 
equilateral antenna triangles in HERA.
The blue curves are the closure phases for just the HI 21cm 
signal.  The expected thermal noise in each spectral channel based 
on the signal-to-noise for the HI signal is $\sigma \sim 17^o$,
after redundancy is applied (Wrobel \& Walker 1999).The green curve shows 
the values for the continuum foregrounds only, while the
red curves show the values for the sum of the continuum and HI line signal. Note
that, since the continuum foregrounds dominate the signal by a factor
of $> 10^3$, the green and red curves are indistiguishable on this scale.} 
\end{figure}

\subsection{Adding the Continuum}

Figure 1 also includes the closure phase vs. frequency curves for the
continuum model, and the continuum plus line models.  The frequency
structure for the continuum emission is much smoother than that seen
for the line signal, although not completely structure-less.  Note
that on the baselines in question (6 wavelengths to 20 wavelengths =
14.6m to 44m), the continuum foreground visibility amplitudes are
$\sim 10$Jy to 60Jy, while the HI 21cm line visibilities are $\sim
4$mJy. Hence, the continuum foregrounds dominate the signal by a
factor of $\sim 10^3$ to $10^4$ (in terms of Jansky per visibility),
and the green and red curves are indistiguishable on this scale.

Fig 2 shows the continuum closure spectra for the three triangles. In
this case, a mean value has been removed, in order to enhance the
scale. The shortest triangle shows the most frequency structure, while
the longer triangles show smoother behaviour of closure phase with
frequency.

\begin{figure}[ht]
\epsscale{0.6}
\plotone{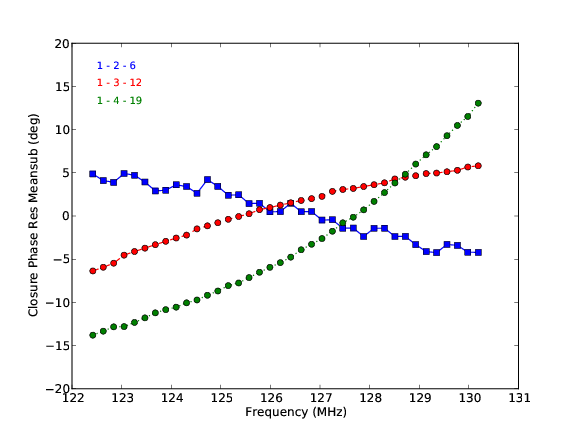}
\caption{\small
The closure phase spectra of the continuum foregrounds only, now with 
a mean value subtracted to enhance scale. 
}
\end{figure}

The one hopeful result is that the continuum closure spectra appear
much smoother with frequency than the line spectra. But are they
smooth enough to remove a smooth curve in frequency in order to
recover the line signature?

We note that we cannot rule-out the possibility that the closure phase
structure of the continuum emission is an artefact of our sky models
and mock-observation process. However, that would be wishful thinking.

\subsection{Line plus continuum with a smooth spectral model removed}

We consider the possibility of removing the continuum via low order
polynomial fitting in frequency.  Fig 3 shows the results of the
closure spectra for a summed line plus continuum model, but with a
third order polynomial fit in frequency and subtracted. The residuals
are then multiplied by 100.  Also included are the HI-only closure
spectra.  The idea is to look for structure in the residual
corresponding to the contribution of the line signal.

We have performed this analysis for all three closure triangle
lengths. The shortest closure triangle shows the largest frequency
dependent structure of the continuum. This may be due to 'small number
statistics', ie. that on the shortest baselines we only have a few
independent resolution elements over the primary beam.  The residuals
become smaller on longer baselines, due to the iherently smoother
continuum closure spectra. However, comparing to the expected HI
closure spectra, there is still no sign of the HI signal.

Keep in mind that polynomial fitting does not take into consideration
possible covariance between the 21-cm signal and the spectral
structure in the foreground model, and can potentially also remove
some of the HI signal. As such, simultaneous estimation of the 21-cm
closure signal and our foreground model would be much more effective
for recovering unbiased estimates, as has been shown in the general
Bayesain approach to HI 21cm power spectral estimation by Sims et
al. (2016) and Lentati et al. (2016).

\begin{figure}[ht]
\epsscale{0.47}
\plotone{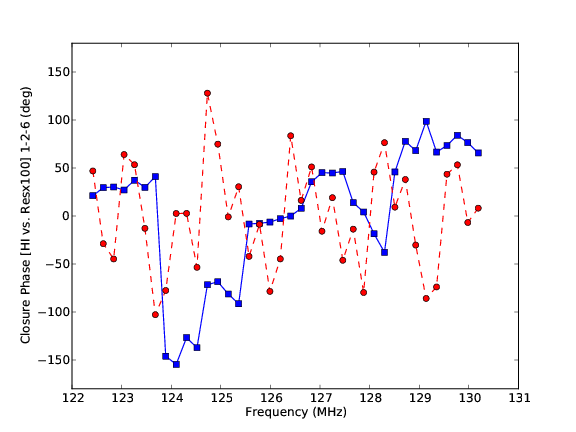}
\plotone{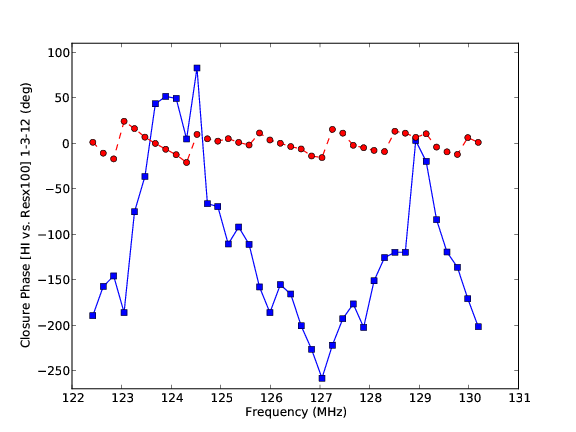}
\plotone{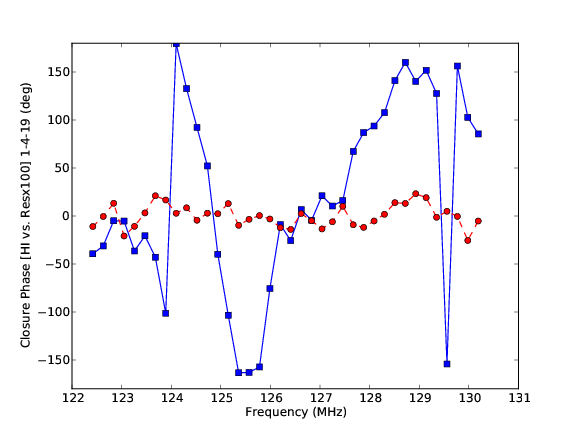}
\caption{\small
The closure phase spectra for the three triangles. The blue curve
shows the HI line spectra only. The red curve shows the continuum plus
redshifted 21-cm line closure spectra residuals following fitting and
subtraction of a third order polynomial. These residuals are them
multiplied by 100.
}
\end{figure}

\subsection{Removal of 99\% of the continuum}

Lastly, we ask the question: what fraction of the continuum needs to
be removed in order to recover the line signal? Figure 4 shows the
results assuming 99\% of the continuum can be removed via imaging and
subtraction of the resulting sky model from the visibilities as a
function of frequency. We only consider one triangle, as
representative. 

The red curve shows the residual closure spectrum after continuum
subtraction from the visibilities, and subsequent third order
polynomial fitting and removal in frequency for the closure spectrum.
The residuals are then multiplied by 30. Also shown is the HI-only
spectrum in blue.

In this case, there is an apparent  correlation between the expected HI
signal, and the residuals after polynomial fitting. The broad
structures are generally reproduced, with some gradual deviation due
to the assumed 3rd order fit.

We should point out that such a process may obviate the need for a
closure analysis, since it assumes that excellent calibration as a
funciton of frequency will be obtained to generate the model images
for subtraction.

\begin{figure}[ht]
\epsscale{0.7}
\plotone{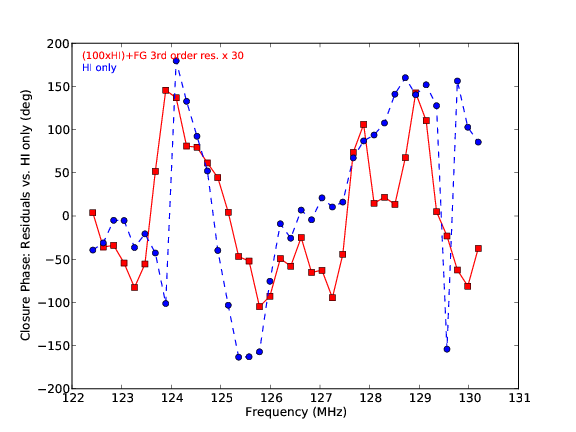}
\caption{\small
Closure phase spectrum for the 1 -- 4 -- 19 triangle. The blue curve
shows the HI line closure spectrum only. The red curve shows the closure
spectrum for a model in which
99\% of the continuum has been subtracted cleanly 
from the continuum plus HI 21cm line visibility data. 
Then a third order polynomial is fit to the closure spectrum
and removed, and the residuals multiplied by 30. 
}
\end{figure}

\section{Conclusions}

We have considered closure phase in the context of detecting the HI
21cm signal from the neutral IGM. The good news is that the closure
spectra for the HI signal show dramatic structure in frequency, and
based on thermal noise alone, the redundant HERA-331 array should
detect these fluctuations easily.  Also good news is that the
frequency structure in the continuum closure spectra is much smoother
than that seen in the HI closure spectra.

The bad news remains the extreme brightness difference between the
continuum and line emission.  The continuum dominates the visibilities
at the level of $10^3$ to $10^4$. We have investigated smooth-curve
fitting and removal to the line plus continuum closure spectra, and
find that the continuum itself shows enough structure in frequency in
the closure spectra to preclude separation of the continuum and line
based on such smooth-curve fitting and removal.

We have also considered the possibility of subtraction of the
continuum using a sky model (derived from the observations), prior to
calculation of the closure spectra. We find that if 99\% of the
continuum can be subtracted from the visibilities prior to
calculation of closure spectra, then the line signal can be seen in
the residuals after subsequent smooth curve fitting and removal. Of
course, if we can calibrate the data to the level required for sky
model generation and subtraction, the need for a closure analysis
becomes potentially redundant.

We can speculate that the natural chromatic response of the
interferometer is causing, in part, the spectral structure in the
closue spectra of the continuum. In this case, there may be a
'wedge-like' approach that could be used, in which the statistics of
the frequency-dependent structure in the closure spectra (eg. a
closure power spectrum), is signficantly different for the line
vs. the continuum, thereby allowing for isolation of the line signal
in some power-spectral domain.  

Similarly, we will explore a Bayesian approach to closure phase
analysis, in which any covariance between the foregrounds and the
21-cm signal in the data is dealt with explicitly through joint
estimation of the foregrounds and 21-cm signatures.  The advantage of
employing closure quantities over the current visibility-based
analyses remains the robustness of the measured closure quantities to
calibration and calibration errors.

\vskip 0.2in

{\bf References} 

Jennison 1958, MNRAS, 118, 276

Carilli, C. \& Sims, P. 2016, HERA Memo. No. 12

Cornwell \& Fomalont 1999, in Synthesis Imaging in Radio Astronomy II,' 
eds. G. B. Taylor, C. L. Carilli, and R. A. Perley. ASP 180, 1999, p. 187

de Boer, D. et al. 2016, PASP, submitted 

Datta, A., Bowman, J. , Carilli, C. 2010, ApJ, 724, 526 

Datta, A., Bhatnagar, S. , Carilli, C. 2009, ApJ, 703, 185 

Dillon, J., \& Parsons, A. 2016, ApJ, submitted (arXiv:1602.06259

Lentati, L. et al. 2016, MNRAS, submitted

Mesinger, A. \& Furlanetto, S., Cen, R. 2011, MNRAS, 411, 955

Morales, M. et al. 2012, ApJ, 752, 137

Morales, K. \& Wyithe, S. 2010, ARAA, 48, 127

Parsons, A. et al. 2012, apJ, 756, 165

Perley, R. \& Fomalont, E. 1999, in Synthesis Imaging in Radio Astronomy II,' 
eds. G. B. Taylor, C. L. Carilli, and R. A. Perley. ASP 180, 1999, p. 79

Perley, R. 1999, in Synthesis Imaging in Radio Astronomy II,'
eds. G. B. Taylor, C. L. Carilli, and R. A. Perley. ASP 180, 1999,
p. 275

Pober, J. et al. 2016, ApJ, 819, 8

Sims, P. et al. 2016, MNRAS, submitted. 

Thompson, A.R., Moran, J., Swenson, G. 2007, Interferometry and
Synthesis in Radio Astronomy, John Wiley \& Sons, 2007.

Wrobel, J. \& Walker, R.C. 1999, in Synthesis Imaging in Radio Astronomy II,'
eds. G. B. Taylor, C. L. Carilli, and R. A. Perley. ASP 180, 1999,
p. 171

\end{document}